\documentclass[aps,prb,twocolumn,floatfix,superscriptaddress,longbibliography]{revtex4-2}

\pdfoutput=1
\usepackage{graphicx}
\usepackage{amssymb}
\usepackage{amsfonts}
\usepackage{amsmath}
\usepackage{hyperref}
\usepackage{mathrsfs}
\usepackage{color}
\usepackage{subfig}
\usepackage{caption}
\newcommand{\bsub}{\begin{subequations}}
\newcommand{\esub}{\end{subequations}}

\newcommand \bea {\begin{eqnarray} }
\newcommand \eea {\end{eqnarray}}

\newcommand{\beg}{\begin{equation}}
\newcommand{\en}{\end{equation}}

\newcommand \bel  {\begin{eqnarray}\begin{aligned}}
\newcommand \enl  {\end{aligned}\end{eqnarray}}

\newcommand{\bp}{\mathbf p}

\newcommand{\bk}{\mathbf k}

\newcommand{\bR}{\mathbf R}
\newcommand{\bS}{\mathbf S}
\newcommand{\bB}{\mathbf B}

\newcommand{\abs}[1]{\lvert #1\rvert}
\newcommand{\eps}{\varepsilon}

\newcommand{\dg}{^\dagger}
\newcommand{\st}{^\ast}

\newcommand{\pmat}{\begin{pmatrix}}
\newcommand{\epmat}{\end{pmatrix}}

\def\v{{\bf v}}
\def\8{\infty}
\def\oh{\frac{1}{2}}

\def\tt{\frac{2}{3}}

\def\sq2{\sqrt{2}}
\def\d{\partial}

\def\undertext#1{\vtop{\hbox{#1}\kern 1pt \hrule}}

\def\be{\begin{equation}}
\def\ee{\end{equation}}
\def\bea{\begin{eqnarray} & &}
\def\eea{\end{eqnarray}}

\def\r{{\bf r}}

\def\b{{\hat b}}

\begin{document}
\title{Magnetic field control of topological magnon-polaron bands in two-dimensional ferromagnets}

\author{Pengtao Shen and Se Kwon Kim\\
\it{Department of Physics and Astronomy, University of Missouri, Columbia, Missouri 65211, USA}}

\date{\today}

\begin{abstract}
	We theoretically study magnon-phonon hybrid excitations in a square lattice ferromagnet subjected to a magnetic field by varying the field direction. The bulk bands of hybrid excitations, which are referred to as magnon-polarons, are investigated by considering all three phonon modes: vertical phonon, transverse phonon, and longitudinal phonon. We show that the topological proprieties of three hybridizations are different in terms of the Berry curvature and the Chern numbers. We also find that the topological properties of the bands can be controlled by changing the direction of the  magnetic field, exhibiting one or more topological phase transitions. The dependence of thermal Hall conductivity as a function of magnetic field direction is proposed as an experiment probe of our theoretical results.
	
\end{abstract} 
\maketitle

\section{Introduction}
Since the discovery of the quantum Hall effects~\cite{klitzing1980new,laughlin1981quantized,haldane1988model}, intrinsic topological properties of electronic bands have been a subject of long standing interest. The Berry phase and Berry curvature~\cite{berry1984quantal} of electron bands, which characterize their topological properties, are responsible for various electron transport phenomena such as the anomalous Hall effect~\cite{onoda2006intrinsic,sinitsyn2007anomalous} and the spin Hall effect~\cite{murakami2003dissipationless,kane2005z,bernevig2006quantum}.
The study of the Berry curvature has been extended to collective bosonic excitations such as magnons~\cite{katsura2010theory,onose2010observation,matsumoto2011theoretical,matsumoto2011rotational,shindou2013topological} and phonons~\cite{zhang2010topological,wang2015topological,mousavi2015topologically}. The finite Berry curvature of the bosonic bands can also give rise to various Hall effects such as the thermal Hall effect\cite{katsura2010theory,onose2010observation,matsumoto2011theoretical,matsumoto2011rotational,shindou2013topological,mousavi2015topologically}. Besides these, the hybridized excitation of magnons and phonons, which is called magnetoelastic wave~\cite{ogawa2015photodrive} or magnon-polaron~\cite{kikkawa2016magnon}, has also been predicted to exhibit a finite Berry curvature and possess the associated nontrivial topology, e.g., in ferromagnets with long-range dipolar interaction~\cite{takahashi2016berry} or Dzyaloshinskii-Moriya~(DM) interaction~\cite{zhang2019thermal}, in noncollinear antiferromagnets with exchange magnetorestriction~\cite{park2019topological}, and in collinear ferrimagnets with DM-induced magnon-phonon coupling~\cite{park2019ferri}. 

Even without such long-range dipolar interaction, DM interaction, or special lattice symmetry, recent works~\cite{go2019topological,zhang2019} have shown that the nontrivial topology of magnon-polaron can emerge in a two-dimensional (2D) magnet by taking account of the well-known magnetoelastic interaction driven by Kittel~\cite{kittel1958magph}. In particular, it has been shown that the topological structure can be controlled by the magnetic field via the change of the number of band-crossing lines. However, the previous investigations have been restricted to the cases where the magnetic field and the ground-state spin direction are perpendicular to the plane, and, in that case, magnons only couple with out-of-plane (ZA) phonons, not affecting the other in-plane phonon modes.

\begin{figure}[t]
	\centering
	\subfloat[]{\includegraphics[width=0.5\linewidth]{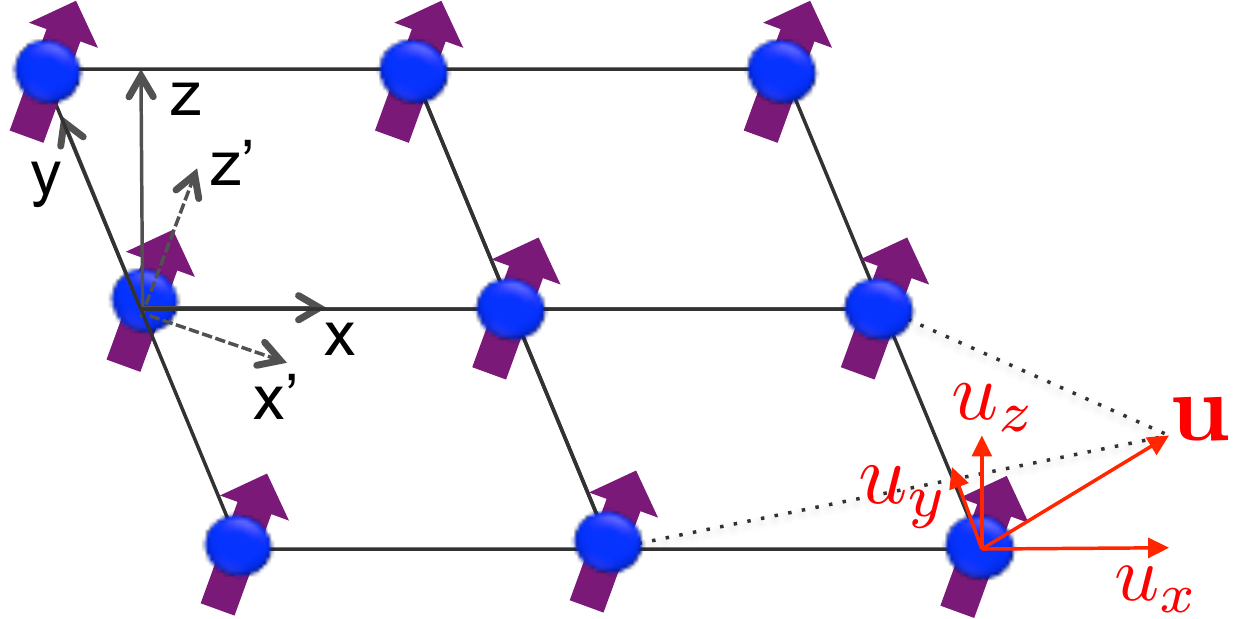}}
    \subfloat[]{\includegraphics[width=0.45\linewidth]{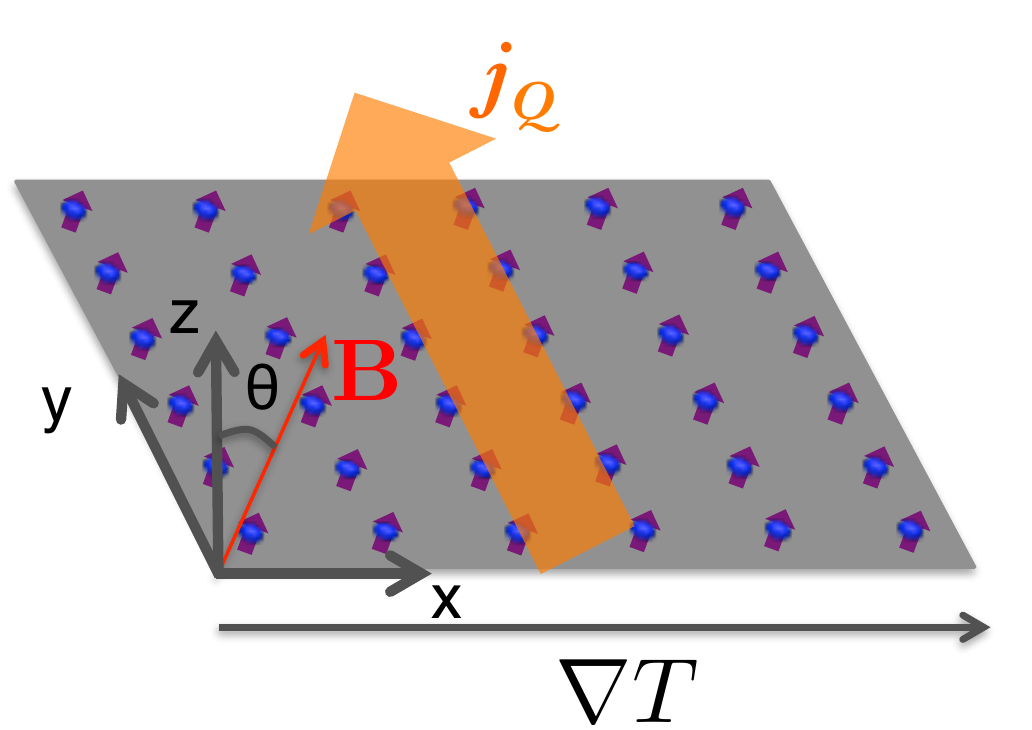}}\\
\caption{(a) The schematic illustration of the model system, where spins denoted by arrows are arranged in a square lattice. The external magnetic field $\mathbf B$ is in $x$-$z$\\ plane with tilt angle $\theta$ from the $z$ axis. The ground state is given by the uniform spin state along the magnetic field $\mathbf{B}$. $x'$-$y$-$z'$ is the rotated coordinate frame, where $z'$ is the spin direction in the ground state. The displacement vector of ions is denoted by the three-dimensional vector $\mathbf u = (u_x, u_y, u_z)$. (b) The schematic illustration of the thermal Hall effect, which refers to the generation of the transverse heat flux $\mathbf{j}_Q$ by a temperature gradient $\nabla T$ in the longitudinal direction. In our work, the thermal Hall effect is induced by the finite Berry curvature of magnon-polarons.}	
\label{schmematic}
\end{figure}

In this paper, we study a more general problem: topological properties of magnon-phonon modes in a square lattice ferromagnet subjected to a magnetic field in an arbitrary direction. See Fig.~\ref{schmematic} for the schematic illustration of the system.  When the magnetic field is away from $z$ direction, the magnons couples not only with ZA phonons, but also with in-plane phonons. Since there are two types of in-plane phonons, longitudinal (LA)  phonon and  transverse (TA) phonon ~\cite{kittel1976introduction,mohr2007phonon,molina2011phonons,nika2012two} in 2D thin film, the dependence of the topological property as well as the thermal Hall conductivity on the field direction are investigated by considering all three phonon modes: vertical, transverse, and longitudinal phonons. Specifically, we investigate the topological structures of the magnon-polaron bands by mapping our model to the well-known two-band model of topological insulator~\cite{bernevig2006quantum} by focusing on each band crossing lines. As an experimental probe, we propose the dependence of thermal Hall conductivity on the magnetic field direction.

Our paper is organized as follows. In Sec.~II, we describe our model including magnetic interaction, elastic interaction and magnetoelastic coupling. In Sec.~III, we study the effect of magnon-phonon coupling on the collective excitations and their topological structures. In Sec IV, we present our results on experimental prediction of the thermal Hall conductivity as a function of magnetic field direction.  We conclude the paper in Sec.~V by providing summary and outlook.

\section{Model}
Our model system is a 2D ferromagnet on a square
lattice described by the Hamiltonian
\bea
H=H_{\text{mag}}+H_{\text{ph}}+H_{\text{mp}} \, ,
\eea
where $H_{\text{mag}}$ and $H_{\text{ph}}$ are the magnetic and elastic subsystem, respectively, and $H_{\text{mp}}$ is the magnetoelastic term between them. We will describe each term in detail below.

\subsection{Magnetic Subsystem}
The magnetic term is given by
\bea
H_{\text{mag}}=-J\sum_{i,j}\bS_i\cdot \bS_j-{\bB}\cdot\sum_i\bS_i \,
\eea
where $J>0$ is the ferromagnetic Heisenberg exchange interaction and $\bB=B(\sin\theta, 0, \cos\theta)$ is the external magnetic field with a tilt angle $\theta$ from z axis rotated about y axis. The ground state of spin direction is along $\bB$, ${\mathbf n}_0=(\sin\theta, 0, \cos\theta)$.

To obtain the magnon band, we will work in a rotated spin frame, where $z'$ is in the direction of $\bB$, $y'$ is the same as y and $x'$ is chosen according to right-handed coordinate system. We perform Holstein-Primakoff transformation,  $S^{x'}_i\approx(\sqrt{2S}/2)(a_i+a\dg_i)$, $S^{y'}_i\approx(\sqrt{2S}/2)(a_i-a\dg_i)$, $S^{z'}_i=S-a\dg_ia_i$,  where $a_i$ and $a\dg_i$ are, respectively, the annihilation and the creation operators of a magnon at site i. By taking the Fourier transformation, $a_i=\sum_k e^{i\bk\cdot \bR_i}a_k/N$, where $N$ is the number of sites in the system, we diagonalize the magnetic Hamiltonian in the momentum space:
\bea\label{magnon}
H_{\text{mag}}=\sum_\bk \hbar \omega_\text{m}(\bk)a\dg_\bk a_\bk \, ,
\eea
where the magnon dispersion is given by $\omega_\text{m}(\bk)=[2JS(2-\cos k_x-\cos k_y)+B]/\hbar$. Here and thereafter, we set lattice constant $a=1$. In long wavelength limit, we obtain the dispersion
\bea
\omega_\text{m}(\bk)=( JSk^2+B)/\hbar\, .
\eea 
The magnon dispersion is quadratic at small wavevector with a gap  $\propto B$.

\subsection{Elastic Subsystem}
The phonon system accounting for the elastic degree of freedom of the lattice is described by the following Hamiltonian:
\bea
H_{\text{ph}}=\sum_i \frac{\bp^2_i}{2M}+\oh\sum_{i,j,\alpha,\beta}u^\alpha_i\Phi^{\alpha,\beta}_{i,j}u^\beta_j \, ,
\eea
where ${\mathbf u}_i$ is the displacement vector of the $i$th ion from its equilibrium position, $\bp_i$
is the conjugate momentum vector, M is the ion mass, and $\Phi^{\alpha,\beta}_{i,j}$ is a force constant matrix between site $i$ and site $j$ ($\alpha,\beta=x,y,z$). We consider a lattice model with in-plane force constant between nearest and second nearest neighbor to be $a_1$ and $a_2$, and out-of-plane force constant between nearest neighbor to be $a_3$. It is convenient to describe in the momentum space, 
\bel
H^{xy}_{\text{ph}}&=\sum_\bk \frac{p^\alpha_{-\bk}{p^\alpha_\bk}}{2M}+\oh { u^\alpha_{-\bk}\Phi^{\alpha\beta}(\bk){u^\beta_\bk}}\,,\\
H^z_{\text{ph}}&=\sum_\bk \frac{p^z_{-\bk}{p^z_\bk}}{2M}+\oh {u^z_{-\bk}\Phi^z(\bk){u^z_\bk}},
\enl
where $\Phi^{xx}(\bk)=2a_1(1-\cos k_x)+2a_2(1-\cos k_x\cos k_y)$, $\Phi^{xy}(\bk)=2 a_2\sin k_x \sin k_y$, $\Phi^{yx}(\bk)=2 a_2\sin k_x \sin k_y$, $\Phi^{yy}(\bk)= 2a_1(1-\cos k_y)+2a_2(1-\cos k_x\cos k_y)$ and $\Phi^z(\bk)=2 a_3(2- \cos k_x-\cos k_y)$. Note that in-plane and out-of-plane mode do not couple~\cite{mohr2007phonon,molina2011phonons,nika2012two}, as $u^x u^z$ and $ u^yu^z$ terms must vanish under the mirror symmetry for 2D system.

We can obtain diagonalized phonon Hamiltonian in quantized phonon operators, $b_{\bk,i}$ and  $b\dg_{\bk,i}$, for longitudinal acoustic(LA) mode,  in-plane transverse acoustic(TA) mode from first equation, for out-of-plane transverse acoustic(ZA) mode from second equation. 
\bea\label{phonon}
H_{\text{ph}}=\sum_\bk\hbar\omega^i_\text{p}({\bk})(b\dg_{\bk,i} b_{\bk,i}+\oh)\, ,
\eea
where $\omega^L_\text{p}({\bk})$, $\omega^T_\text{p}({\bk})$, $\omega^Z_\text{p}({\bk})$ are phonon dispersions for LA, TA and ZA phonon mode, respectively. $\omega^Z_\text{p}(\bk)=\sqrt{{\Phi^z(\bk)}/{M}}$ and $\omega^{L/T}_\text{p}(\bk)=\sqrt{{\Phi^{L/T}(\bk)}/{M}}$, where $\Phi^{L/T}(\bk)$ are two eigenvalues of $\Phi^{\alpha\beta}(\bk)$.
In the long wavelength limit, 
\bel
\omega^Z_\text{p}({\bk})&=v^Z_\text{p} k\,,\\
\omega^L_\text{p}({\bk})&=v^L_\text{p}(\phi_k)k\,,\\
\omega^T_\text{p}({\bk})&=v^T_\text{p}(\phi_k)k\,,
\enl
where phonon velocities are $v^Z_\text{p}$=$\sqrt {{a_3}/M}$ and
$v^{L/T}_\text{p}(\phi_k)$\\$=[(a_1+2a_2\pm\sqrt{\cos ^2 2 \phi_ka_1^2 +4 \sin^2 2 \phi_ka_2^2})/2M]^{1/2}$, where $+$/$-$ are chosen for LA/TA respectively.
It can be easily found from above equation that phonon velocity of LA phonon is larger than TA phonon. We will consider thin film, where less strain is in out-of plane direction than in-plane direction, and therefore we will assume $v^L_\text{p}>v^T_\text{p}>v^Z_\text{p}$~\cite{mohr2007phonon,molina2011phonons,nika2012two}.

\begin{figure}[t]
	\centering
	\subfloat[]{\includegraphics[width=0.45\linewidth]{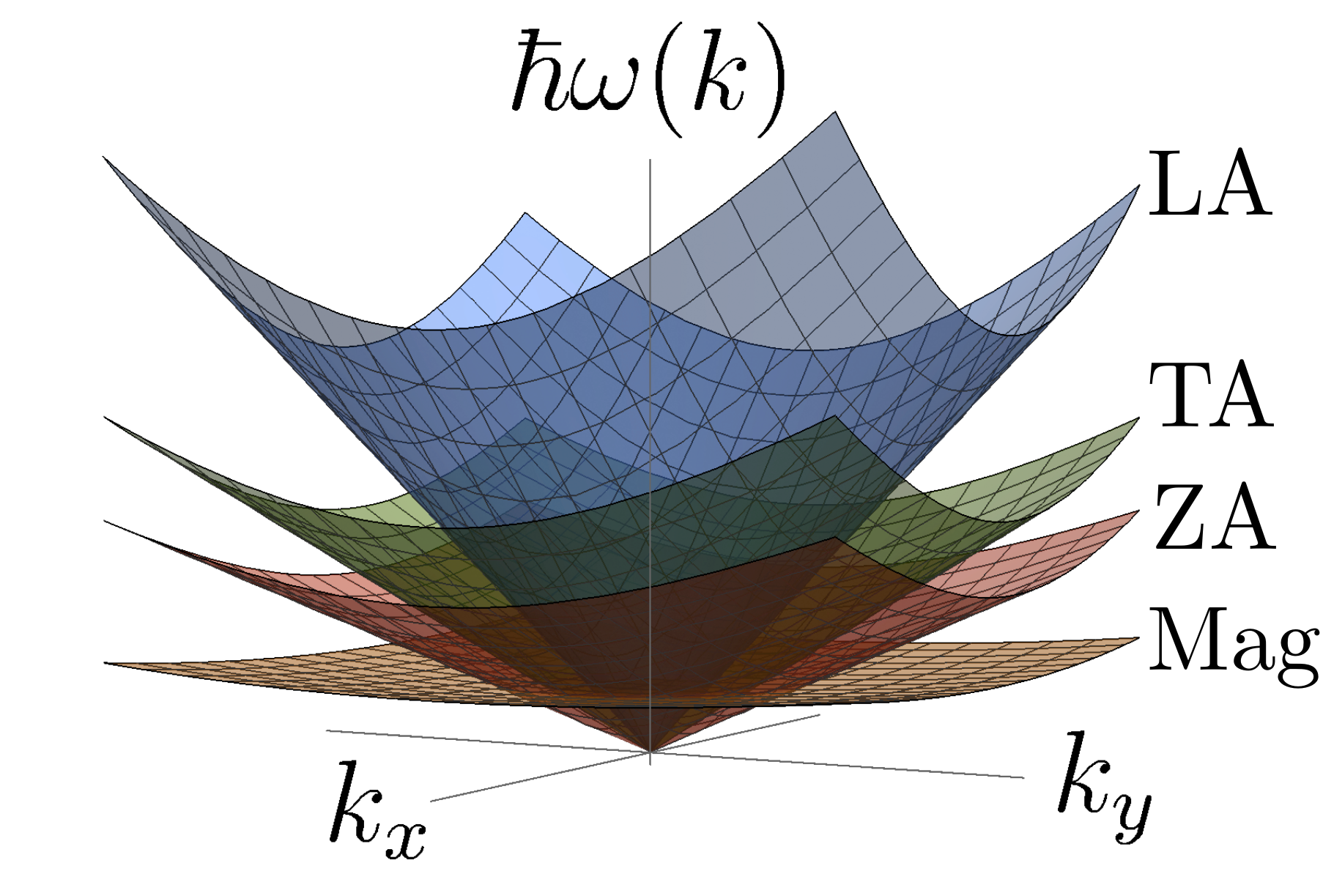}}
	\quad
	\subfloat[]{\includegraphics[width=0.45\linewidth]{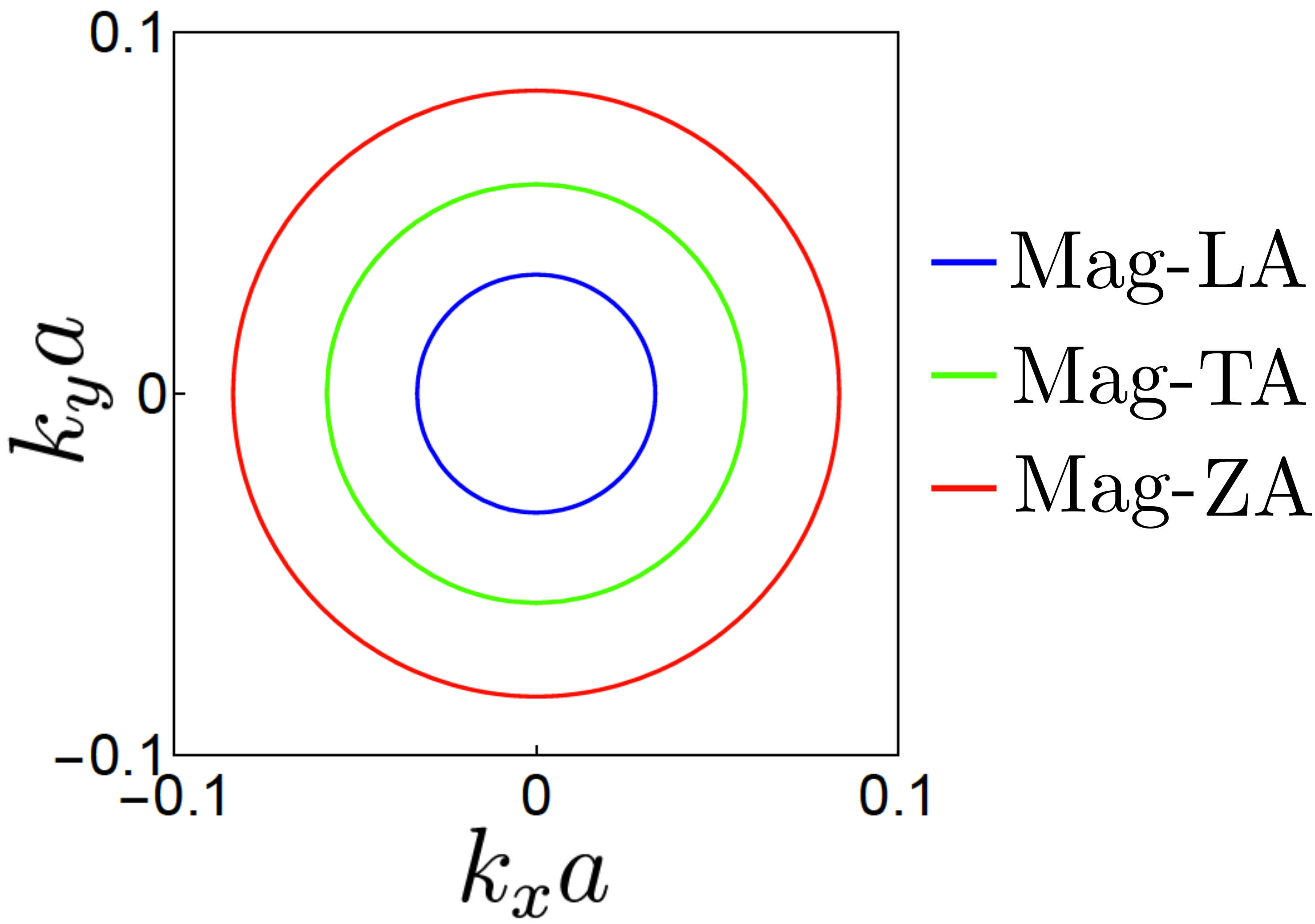}}\\
	\subfloat[]{\includegraphics[width=0.43\linewidth]{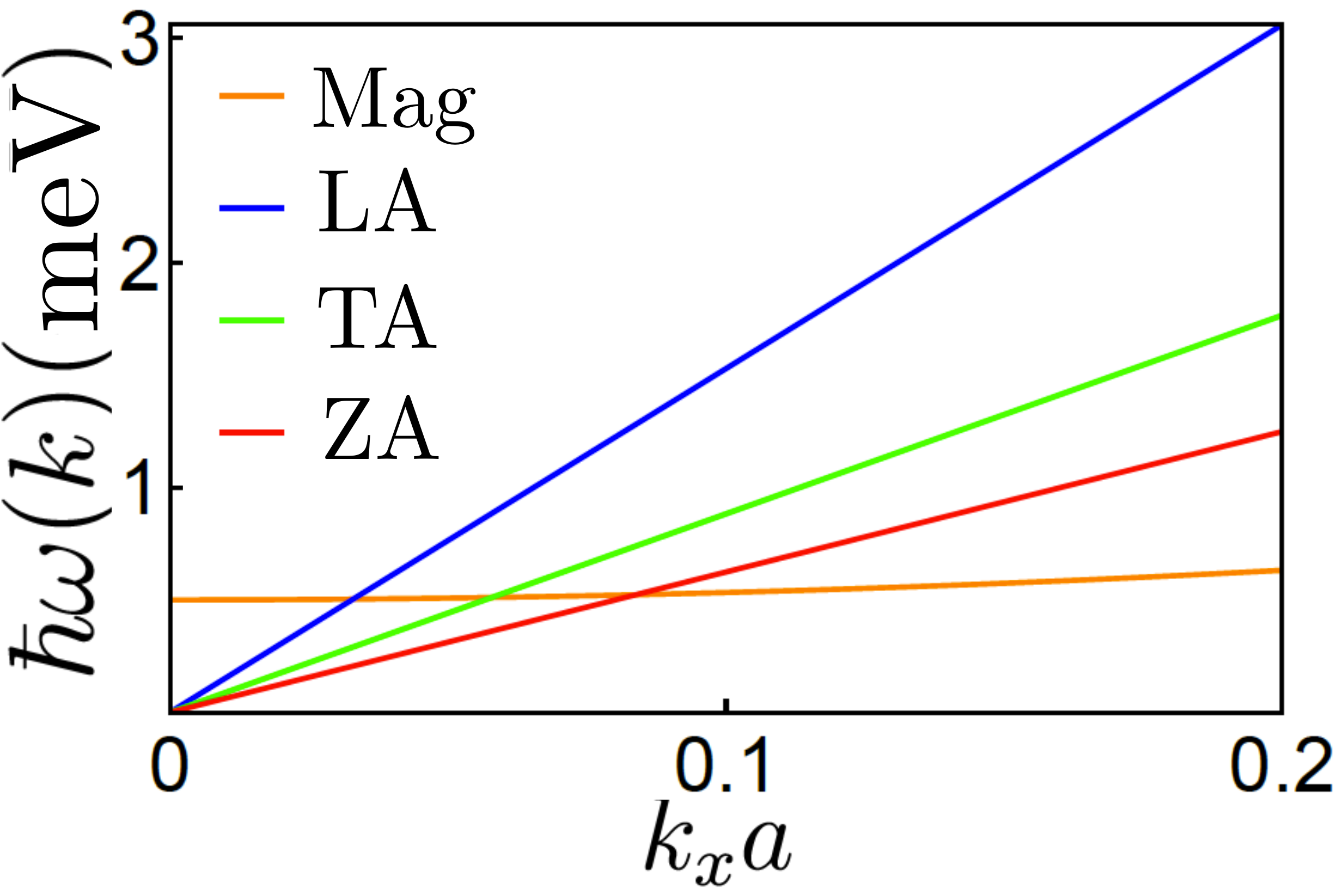}}
    \quad\,\,
	\subfloat[]{\includegraphics[width=0.43\linewidth]{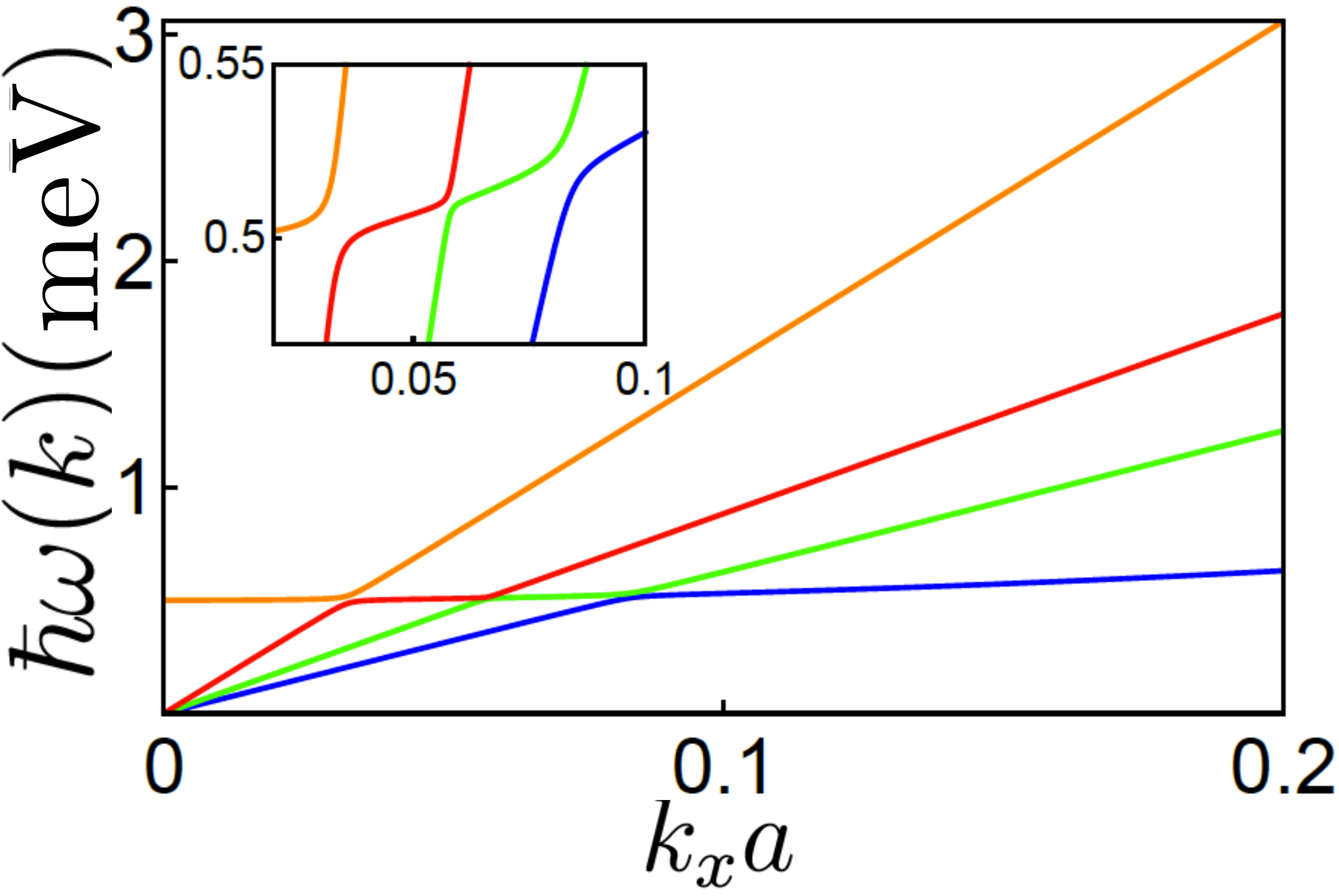}}
	\caption{Dispersion relation of excitation bands 	with magnetic field tilt angle $\theta=\pi/8$ and parameters used are given in the main text of Sec.~III. (a) Dispersion of ZA, LA, TA phonons and magnons, denoted by Mag, without magnetoelastic coupling. (b) Contour plot of the rings of momentum that magnons cross with LA, TA and ZA phonons. (c) Dispersion in $k_x$ direction without magnetoelastic coupling. (d) Hybridization of magnons and ZA, LA, TA phonons in $k_x$ direction in the presence of magnetoelastic coupling. Inset is a zoom-in of magnons and phonons crossing. 
    }
	\label{dispersion}
\end{figure}

The phonon operators are introduced in such a way that
\bel
u^Z_\bk&=\oh\sqrt{\frac{\hbar}{M \omega^Z_\text{p}(\bk)}}\left(\frac{b_{\bk,Z}+b\dg_{-\bk,Z}}{\sq2}\right)\,, \\
p^Z_\bk&=\sqrt{\hbar M \omega^Z_p(\bk)}\left(\frac{b_{-\bk,Z}-b \dg_{\bk,Z}}{\sq2 i}\right) \,,
\enl
and for in plane vibration, 
\bel
{\mathbf u}^{L/T}_\bk&=\sqrt{\frac{\hbar}{M \omega^{L/T}_\text{p}(\bk)}}\left(\frac{b_{\bk,L/T}+b \dg_{-\bk,L/T}}{\sq2}\right) \hat{\mathbf  e}^{L/T}(\bk)\,,\\
{\mathbf p}^{L/T}_\bk&=\sqrt{\hbar M \omega^{L/T}_\text{p}(\bk)}\left(\frac{b_{-\bk,L/T}-b \dg_{\bk,L/T}}{\sq2 i}\right)\hat {\mathbf  e}^{L/T}(\bk)\,,
\enl
where $\hat {\mathbf  e}^{L/T}(\bk)$ are the normalized two-dimensional eigenvectors of $\Phi^{\alpha\beta}(\bk)$, representing the $x,y$ components of the mode. These vectors are given by  $\hat {\mathbf  e}^L(\bk)=(\cos\phi_k,\sin\phi_k)$ and $\hat {\mathbf  e}^T(\bk)=(\sin\phi_k,-\cos\phi_k)$ for a pure longitudinal and transverse mode, but generally speaking, the LA and TA phonon modes are not entirely longitudinal or transverse, $\hat {\mathbf  e}^L(\bk)\nparallel \bk$ and  $\hat {\mathbf  e}^T(\bk)\not\perp\bk$, unless $\bk = (k_x, k_y)$ are along the high symmetry directions, e.g., (1, 0), (1, 1), (0, 1). 
One special case is that $a_1=2a_2$, when $\omega^{L/T}_\text{p}({\bk})$ are isotropic and LA and TA phonons are pure longitudinal and transverse in any direction in long wavelength limit. We show the magnon and three phonon dispersions in Fig.~\ref{dispersion}. Without magnon-phonon coupling, the gappd magnon band crosses with each of the gapless phonon band at a ring of momentum satisfies $\omega_\text{m}(\bk)=\omega^i_\text{p}(\bk)$. We do not include the second possible band crossing at larger momentum~\cite{go2019topological} as they are of much higher energy, and therefore less significant for transport at sufficiently low temperature.

\subsection{Magnetoelastic Coupling}
The magnetoelastic coupling is generally modeled by the following energy density~\cite{kittel1958magph}, which takes into account of the interaction between the magnetization and lattice deformation due to strain.
\bel
\epsilon_{\text{mp}}&=\kappa_1 (S^2_x e_{xx}+S^2_y e_{yy}+S^2_z e_{zz})\\
&+\kappa_2 (S_x S_y e_{xy}+S_x S_z e_{xz}+S_y S_z e_{yz})\,,
\enl
where $e_{mn}={\d u^m}/{\d n}+{\d u^n}/{\d m}$. 

We can use the three-dimensional rotation matrix $\mathcal R$ that transforms the $z$ axis to the equilibrium spin direction ${\mathbf n}_0$: ${\mathbf n}_0=\mathcal R {\mathbf n}'_0$, with ${\mathbf n}'_0=\hat z$ and  ${\mathbf n}_0=(\sin\theta, 0, \cos\theta)$
Using the rotational transformation, 
\bea
\left(
\begin{array}{ccc}
S_x\\S_y\\S_z
\end{array} 
\right)
=
\left(
\begin{array}{ccc}
	\cos \theta   & 0&  \sin \theta  \\
	0 & 1 & 0 \\
	-\sin \theta  & 0 & \cos \theta  \\
\end{array}
\right)
\left(
\begin{array}{ccc}
	S_{x'}\\S_{y'}\\S_{z'}
\end{array} 
\right)\,,
\eea
we obtain the magnetoelastic energy density in linear order of the magnon amplitude, $S_{x'}$ and $S_{y'}$,
\bel\label{kittel}
\epsilon_{\text{mp}}&=\kappa_1 \sin 2\theta S_{x'}S_{z'} e_{xx}+\kappa_2 \sin\theta S_{y'}S_{z'} e_{xy}\\
&+\kappa_2\cos2\theta S_{x'}S_{z'} e_{xz}+\kappa_2\cos\theta S_{y'}S_{z'} e_{yz}
\enl
In terms of the magnon and phonon operator introduced earlier in Eq.~(\ref{magnon}) and Eq.~(\ref{phonon}), the magnetoelastic coupling term can be recast into particle-number-conserving terms, $a\dg b, ab\dg$, and particle-number-nonconserving terms, $a\dg b\dg, ab$. We follow Ref.~\cite{go2019topological} by neglecting the particle-number-non-conserving terms because the effect of those on band structure is of quadratic order in the magnetoelastic coupling and thus is much smaller than the effect of  the particle-number-conserving terms. 

The particle-number-conserving Hamiltonian is then given by
\bea
H\approx\sum_\bk(a\dg_\bk,b^{L\dagger}_\bk,b^{T\dagger}_\bk,b^{Z\dagger}_\bk)\mathcal{H}_k
\left(
\begin{array}{c}
	a_\bk\\b^{L}_\bk\\b^{T}_\bk\\b^{Z}_\bk
\end{array}
\right)\,,
\eea
\begin{figure}[t]
	\centering
	\subfloat[]{\includegraphics[width=0.47\linewidth]{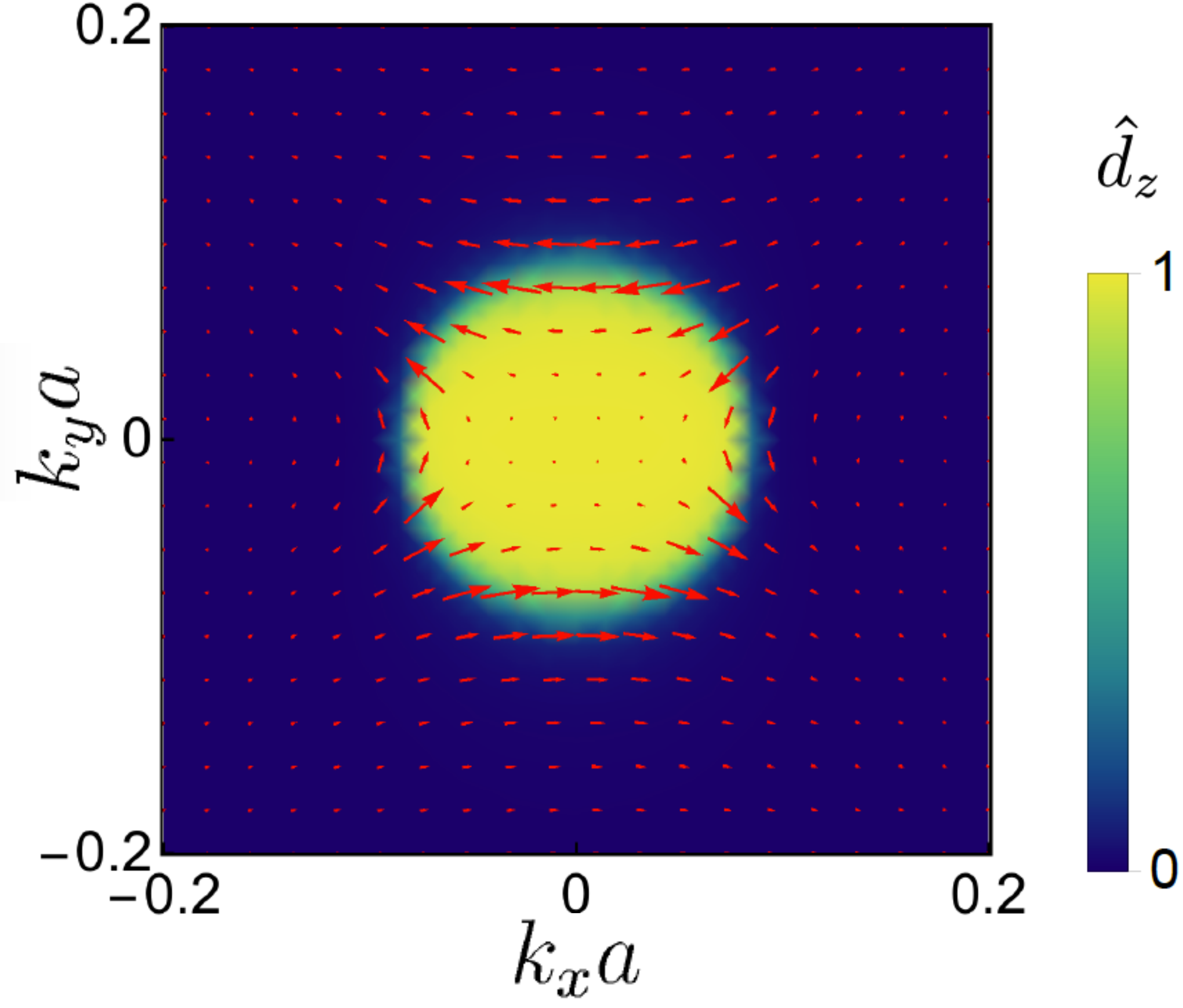}}
	\hfill
	\subfloat[]{\includegraphics[width=0.5\linewidth]{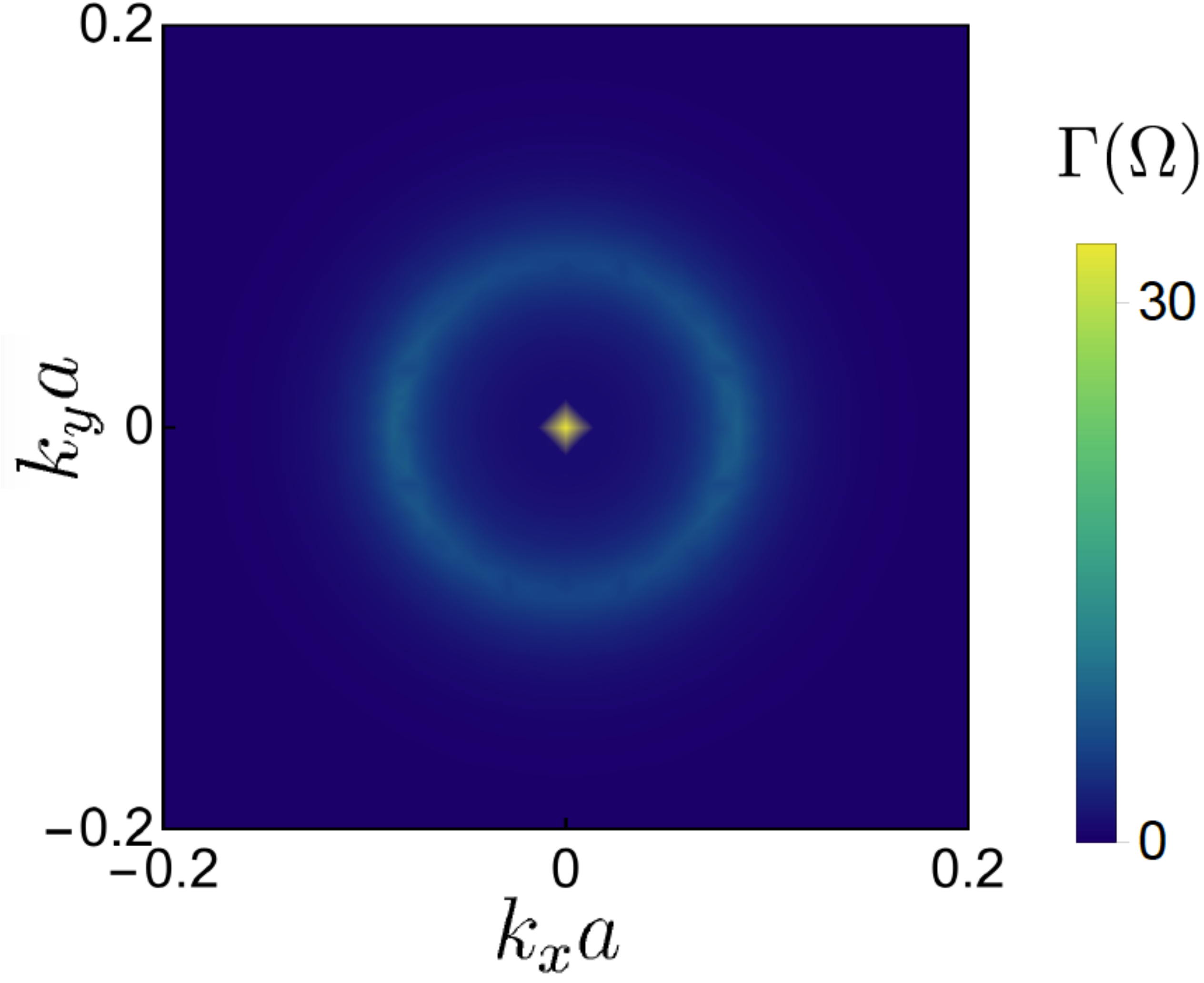}}\\
	\subfloat[]{\includegraphics[width=0.47\linewidth]{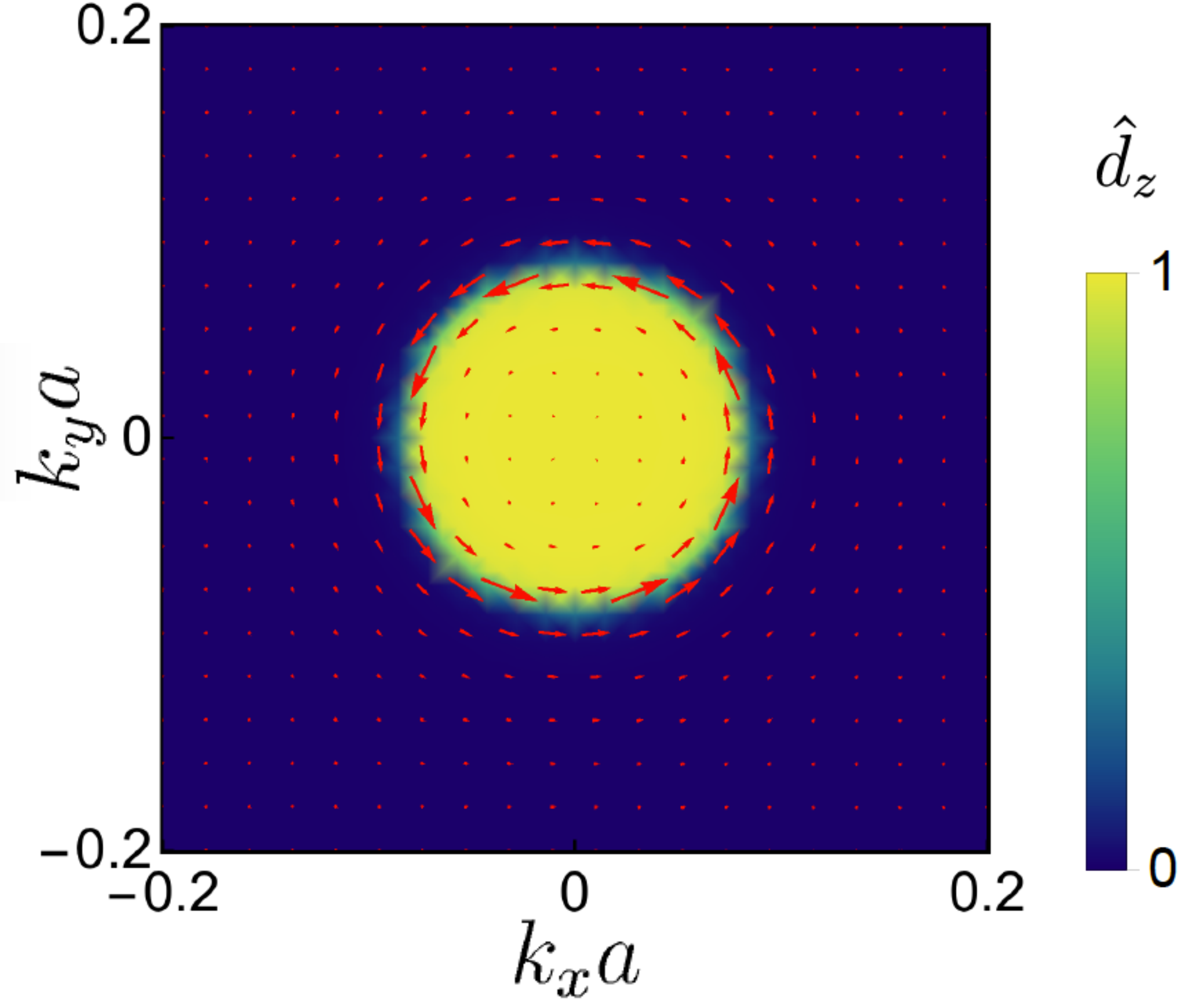}}
	\hfill
	\subfloat[]{\includegraphics[width=0.5\linewidth]{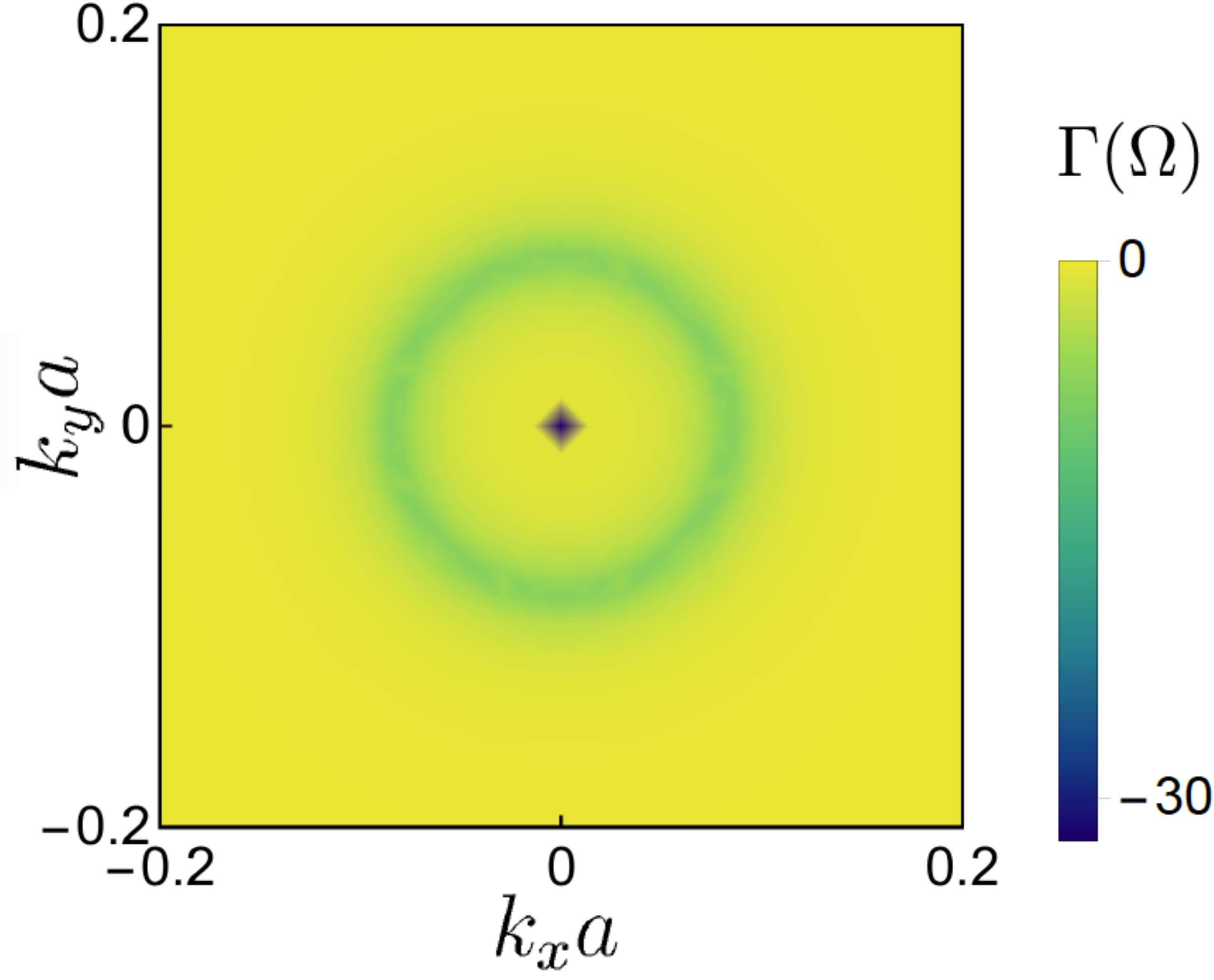}}\\
	\caption{ The Berry curvatures of the upper band in log-scale $\Gamma(\Omega) = \text{sign}(\Omega)\log(1+\abs\Omega)$ (right) and schematic illustration of $\hat{\mathbf d}(\bk)$ (left) from magnon and ZA phonon coupling. The in-plane components of the $\hat{\mathbf{d}}$ vector, $(\hat d_x, \hat d_y)$, are shown as arrows. (a, b) Magnetic field with tilt angle $\theta=\pi/8$. (c, d)  Magnetic field with tilt angle  $\theta=3\pi/8$. }
	\label{ZA}
\end{figure}
with
\bea\label{Hamiltonian}
\mathcal{H}_k=
\left(
\begin{array}{cccc}
		\hbar\omega_m(\bk)&M_L&M_T&M_Z\\
		M\st_L&\hbar\omega^L_\text{p}(\bk)&0&0\\
		M\st_T&0&\hbar\omega^T_\text{p}(\bk)&0\\
		M\st_Z&0&0&\hbar\omega^Z_\text{p}(\bk)\\
\end{array}
\right)\,,
\eea
where
\begin{align}
\nonumber
&M_L=\tilde\kappa_L\left[\frac {\kappa_1} {\kappa_2}2\sin2\theta e^L_xi k_x-\sin\theta e^L_x k_y-\sin\theta e^L_y k_x\right]\,, \\
\nonumber
&M_T=\tilde\kappa_T\left[\frac {\kappa_1} {\kappa_2} 2\sin2\theta e^T_xi k_x-\sin\theta e^T_x k_y-\sin\theta e^T_y k_x\right]\,, \\
&M_Z=\tilde\kappa_Z(\cos2\theta i k_x-\cos\theta k_y)\,,
\end{align}
and $\tilde\kappa_i=\frac{\kappa_2}2 S\sqrt{\frac{\hbar S}{M \omega^i_p(\bk)}}. $

In the case of $a_1=2a_2$ when LA phonon is pure longitudinal and TA phonon is pure transverse,
\bel
M_L&=\tilde\kappa_L\left(\frac {\kappa_1} {\kappa_2} 2\sin2\theta \cos^2 \phi_k i-\sin\theta \sin2\phi_k\right)k\,,\\
M_T&=\tilde\kappa_T\left(\frac {\kappa_1} {\kappa_2} \sin2\theta \sin2\phi_k i+\sin\theta \cos 2\phi_k\right)k\,,\\
M_Z&=\tilde\kappa_Z (\cos2\theta \cos\phi_k i -\cos\theta \sin\phi_k)k\,.
\enl

The band structure of magnon-phonon hybrid system is obtained by diagonalizing Eq.~(\ref{Hamiltonian}). Without magnon-phonon coupling, magnon band crosses with ZA, LA, TA phonon band and the excitations are not hybridized.  With the presence of magnon-phonon couplings, bands open up the gaps at the bands crossing rings and form hybridized excitations, namely magnon-polarons, which induces the nontrivial topological properties of the bands, characterized by the Berry curvature.  See Fig.~\ref{dispersion}(d) for the hybridized bands plot.

\section {topological properties}
The rigorous calculation of band structure and Chern number can be done with SU(4) formalism~\cite{barnett2012su3topo}.  For weak coupling~\cite{zhang2019}, the Berry curvature concentrates in the vicinity of the avoided crossing ring, while the three rings of momentum satisfying $\omega_m(\bk)= \omega^i_p(\bk)$ are relatively separated from each other. We can use the familiar two-band theory of topological insulator~\cite{bernevig2006quantum,qi2008topological} to understand the band topology by focusing on each pair of magnon and phonon bands.

\begin{figure}[t]
	\centering
	\subfloat[]{\includegraphics[width=0.47\linewidth]{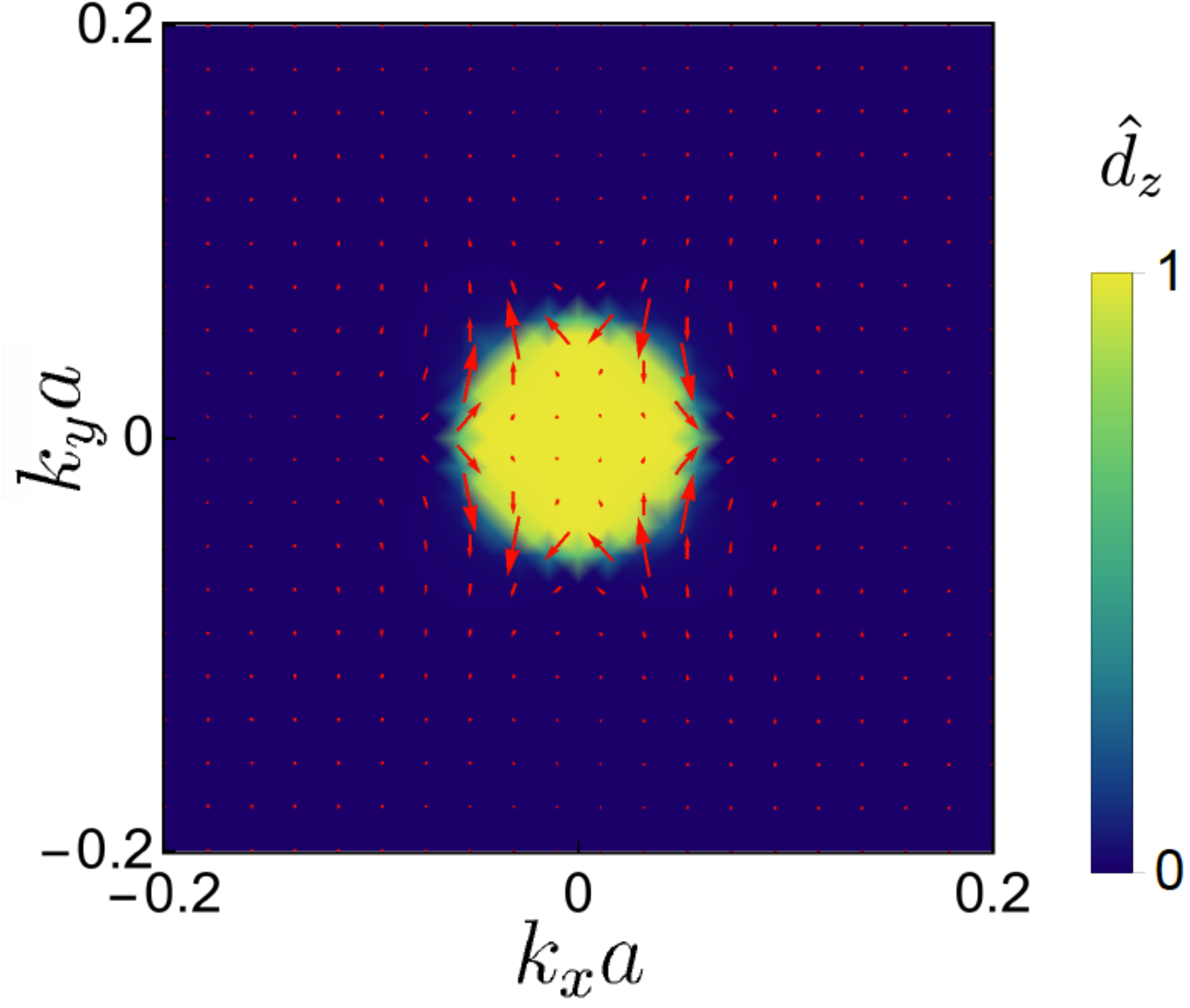}}
	\hfill
	\subfloat[]{\includegraphics[width=0.5\linewidth]{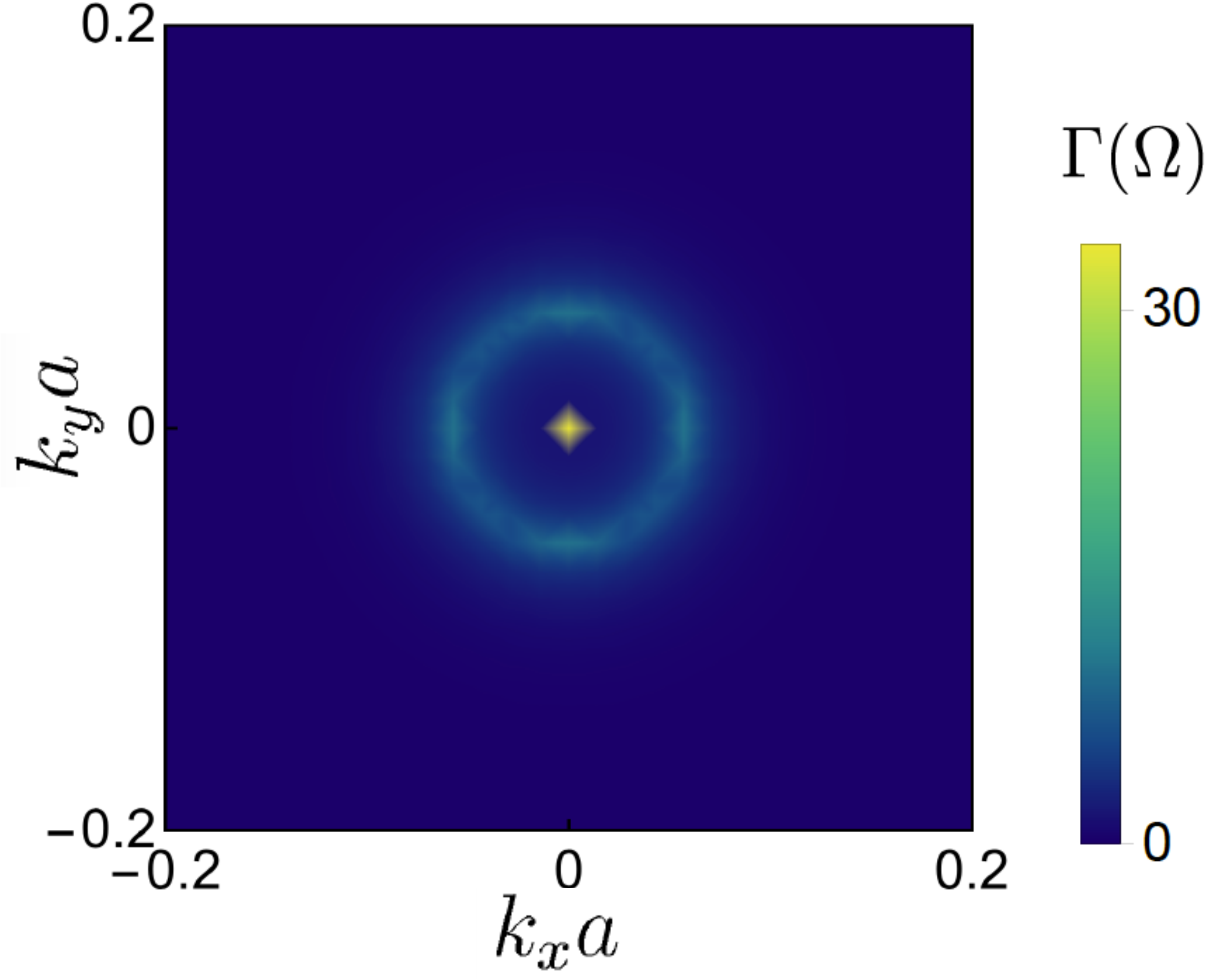}}\\
	\subfloat[]{\includegraphics[width=0.47\linewidth]{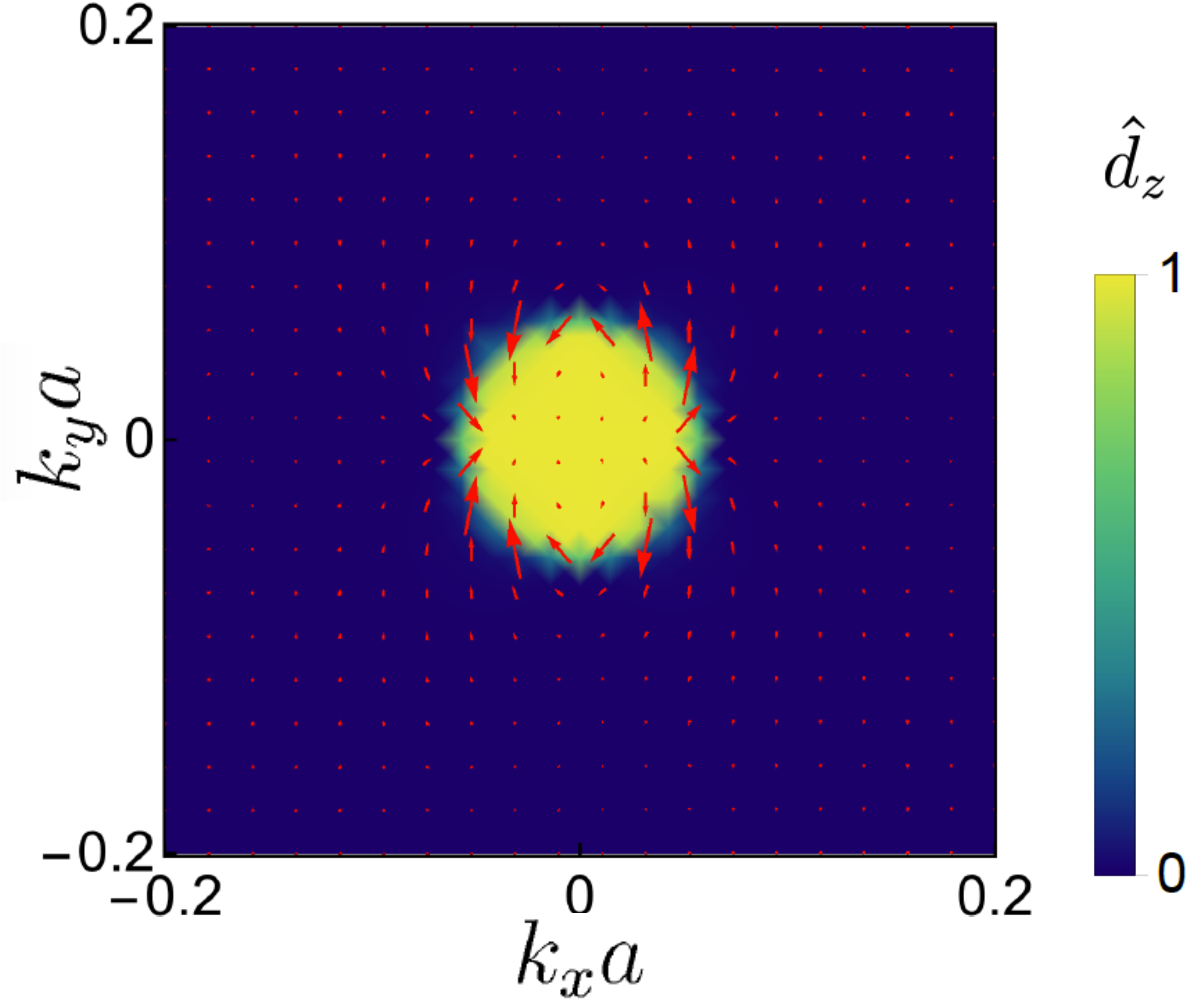}}
	\hfill
	\subfloat[]{\includegraphics[width=0.5\linewidth]{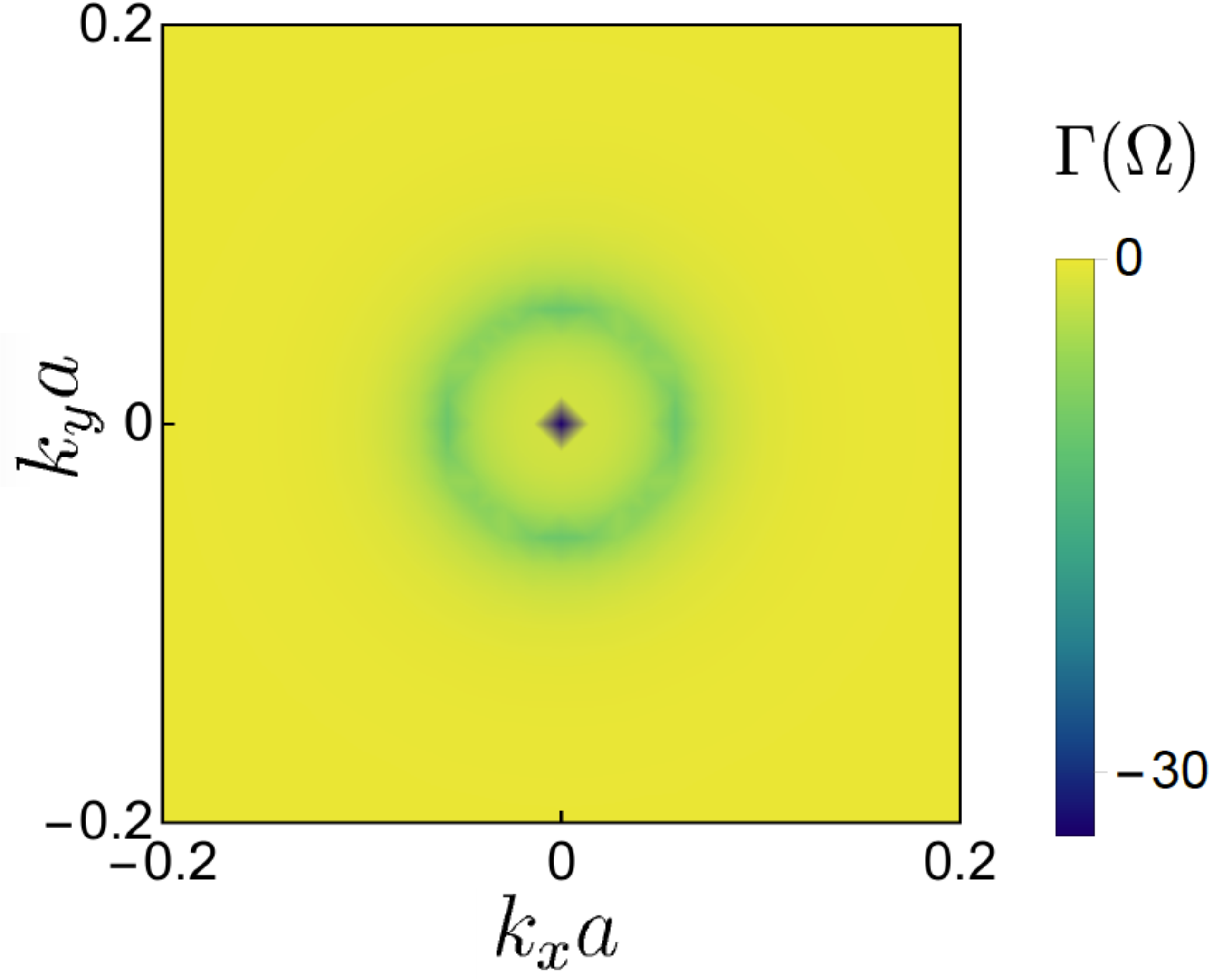}}\\
	\caption{The Berry curvatures of the upper band in log-scale $\Gamma(\Omega) = \text{sign}(\Omega)\log(1+\abs\Omega)$ (right) and schematic illustration of ${\mathbf d}(\bk)$ (left) from magnon and TA phonon coupling, the in-plane components of the $\hat{\mathbf{d}}$ vector, $(\hat d_x, \hat d_y)$ are shown as arrows. (a, b) Magnetic field with tilt angle $\theta=\pi/4$. (c, d) Magnetic field with tilt angle  $\theta=3\pi/4$}
	\label{TA}
\end{figure}

We can write the Bloch Hamiltonian, Eq.~(\ref{Hamiltonian}) in the following form near each band crossing:
\bea
\mathcal{H}_k=
\left(
\begin{array}{ccc}
	\mathcal{H}^i_{2\times 2}(\bk)&0&0\\
	0&\hbar\omega^j_\text{p}(\bk)&0\\
	0&0&\hbar\omega^k_\text{p}(\bk)\\
\end{array}
\right)+\mathcal V_k\,,
\eea
where $\mathcal V_k$ is a perturbation term that does not participate in opening the gap between the two bands ($\eps_\text{m}$, $\eps^i_\text{p}$), and  $\mathcal{H}^i_{2\times 2}(\bk)$ can be written into the form of
\bea\label{2bandH}
\mathcal{H}^i(\bk)=\oh \hbar (\omega_\text{m}+\omega^i_\text{p})I+{\mathbf d}^i(\bk)\cdot \boldsymbol \sigma\,,
\eea 
with $\boldsymbol \sigma=(\sigma_x,\sigma_y,\sigma_z)$ are Pauli matrices and ${\mathbf d}^i(\bk)$ for each band crossing are
\bel\label{d}
{\mathbf d}^Z(\bk)= &[-\tilde\kappa_Z\cos\theta \sin\phi_k k, -\tilde\kappa_Z \cos2\theta \cos \phi_k k,\\ &\oh \hbar (\omega_\text{m}-\omega^Z_\text{p} ) ]\,\\
{\mathbf d}^L(\bk)= & [-\tilde\kappa_L\sin\theta\sin2\phi_k k, -\tilde\kappa_L\frac {\kappa_1} {\kappa_2} 2\sin2\theta \cos^2 \phi_k k,\\ &\oh \hbar (\omega_\text{m}-\omega^L_\text{p} ) ]\,,\\
{\mathbf d}^T(\bk)=& [\tilde\kappa_T\sin\theta \cos2\phi_k k, -\tilde\kappa_T\frac {\kappa_1} {\kappa_2} \sin2\theta \sin2 \phi_k k, \\ &\oh \hbar (\omega_m-\omega^T_\text{p} ) ]\,.
\enl

In terms of  the normalized vector $\hat{\mathbf d}={\mathbf d}/\abs {\mathbf d}$, the Berry curvature is written explicitly as 
\bea
\Omega^i_{\pm}(\bk)=\mp\oh \hat{\mathbf d}^i(\bk)\cdot\left(\frac{\d \hat{\mathbf d}^i(\bk)}{\d k_x}\times\frac{\d \hat{\mathbf d}^i(\bk)}{\d k_y}\right)\,,
\eea
where $+$ and $-$ are for upper and lower bands, respectively.
The corresponding expression for Chern number is given by~\cite{qi2008topological,qi2006topological,volovik1988analog}
\bea
C_{i,\pm}=\frac 1{2\pi}\int d^2k\Omega^i_{\pm}(\bk)\,,
\eea
which is the skyrmion number of the $\hat{\mathbf d}$ vector~\cite{qi2008topological}, counting how many times  $\hat{\mathbf d}$ wraps the unit sphere in the Brillouin zone.
\begin{figure}[t]
	\centering
	\includegraphics[width=1\linewidth]{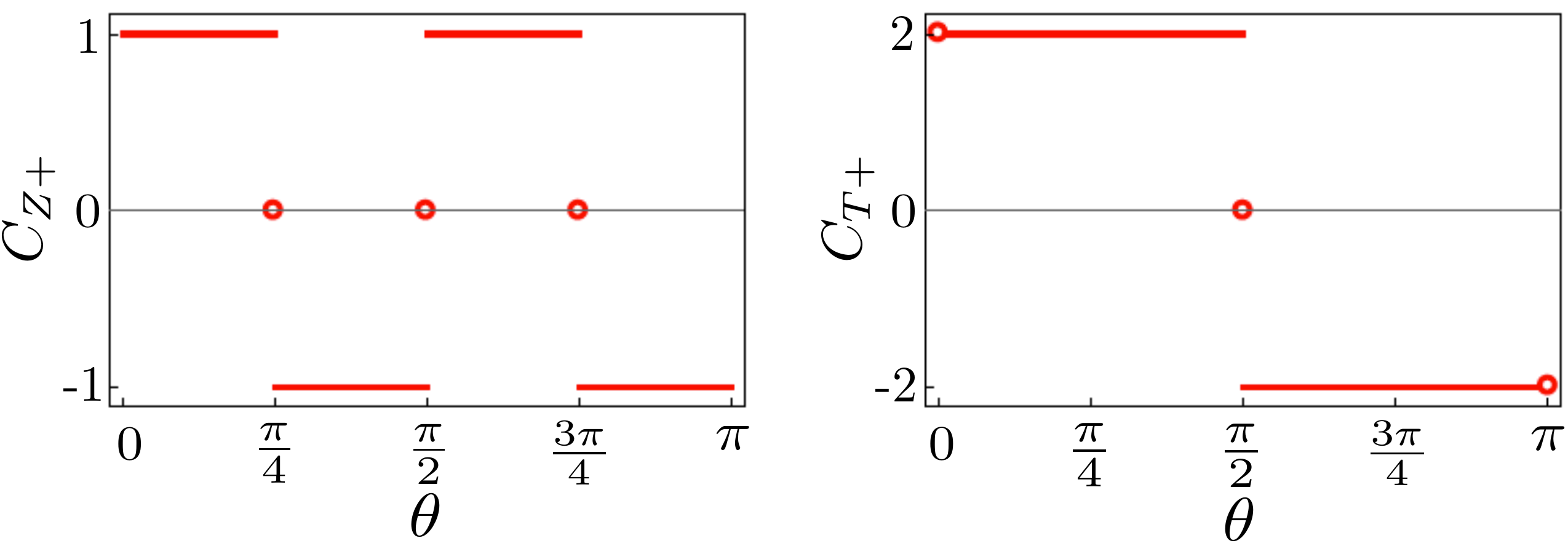}
	\caption{Dependence of the Chern number of  upper branch $C_{i,+}$ on tilt angle $\theta$ (a) for the hybridization of magnons and ZA phonons and (b) for the hybridization of magnons and TA phonons. $\abs{C_{Z,+}}=1$ and its sign changes at $\theta=\pi/4, \pi/2$ and $3\pi/4$.  $\abs{C_{T,+}}=2$ and its its sign changes at $\theta=\pi/2$. }
	\label{C-theta}
\end{figure}

We note that the magnon and LA phonon coupling aways vanishes in $k_y$ direction. This can be understood from Eq.~(\ref{kittel}). As we set the magnetic field in x-z plane, there is no term with $e_{yy}=\d u_y/dy$. This means LA phonons at $k_y$ direction does not couple to magnons and the two hybridized bands are touching at $(0, \pm k_y)$. The Berry curvatures at the touching points are not well defined. Similar phenomena has also been found in Weyl semimetal and such points are called Weyl points~\cite{armitage2018weyl}. The corresponding Chern number from magnon and ZA phonon coupling is $\abs {C_Z}=1$ and one for magnon and TA phonon coupling is $\abs {C_T}=2$, which can be easily understood by the dependence of polar angle $\phi_k$ in Eq.~(\ref{d}). Prefactor of  $d^Z_x$ and $d^T_y$ change sign across $\theta=\pi/2$, and prefactor of $d^Z_y$ changes sign across $\theta=\pi/4$~and~$3\pi/4$. Across each of these angles, $\hat{\mathbf d}$ changes its winding direction, and therefore there is a topological transition as well. See Fig.~\ref{C-theta} for the dependence of Chern number on magnetic filed tilt angle $\theta$ from 0 to $\pi$. When $\theta=\pi/4$ or $3\pi/4$, the mangnon and ZA phonon coupling vanishes in the $k_x$ direction. When $\theta=\pi/2$, the mangnon and ZA phonon coupling vanishes in the $k_y$ direction and the mangnon and TA phonon coupling vanishes in the direction of $\phi_k={n\pi}/4$ with odd integer $n$. The Berry curvatures are not well defined at the band touching points at these critical angles. When $\theta=0$ and~$\pi$, there is no coupling between mangons and LA, TA phonons at any momentum.  Hybridization only occurs between magnons and ZA phonons as previously reported~\cite{go2019topological}.

The above discussion about band touching and topological transition holds even if $a_1\neq2a_2$ since the phonon modes are always longitudinal or transverse in the direction of $\phi_k={n\pi}/4$ with integer $n$. The effect can be shown by replacing $\phi_k$ in $\hat {\mathbf  e}^{L/T}(\phi_k)$ with function $\psi(\phi_k)$, where $\psi(\phi_k)$ is a monotonically increasing function of $\phi_k$ and satisfies $\psi(\phi_k)=\phi_k$ at $\phi_k={n\pi}/4$. 

We show the change of the Berry curvatures and the topological properties for magnon and ZA phonon coupling in Fig.~\ref{ZA} and magnon and TA phonon coupling in Fig.~\ref{TA}. For calculation, we used the parameters of the monolayer ferromagnet CrI$_3$ in Refs.~\cite{zhang2019thermal,zhang2015robust,huang2017layer,lado2017origin} ($J$=$2.2$\,meV, $S$=$3/2$, $Mc^2$=$5\times$$10^{10}$\,eV).  The force constant is taken $a_1$=2$\times$$10^5$\,meV/nm$^2$, $a_2$=$10^5$\,meV/nm$^2$, $a_3$=5$\times$$10^4$\,meV/nm$^2$; the coupling strengths are taken $\kappa_1$=10\,meV/nm and $\kappa_2$=5\,meV/nm. The external magnetic field is taken $B$=0.5\,meV. We found the dominant contribution of the Berry curvature comes from vicinity of the ring where two bands cross and it changes sign across the critical angles as shown in Fig.~\ref{ZA}(b,~d) and Fig.~\ref{TA}(b,~d).  Note that there is singularity of the Berry curvature near $k=0$. This originates from the $1/\sqrt{\omega_\bk}$ dependence of $\tilde\kappa$ in ${\mathbf d}(\bk)$. We verified that its contribution to the Chern number and thermal Hall conductivity is negligible despite the divergence. In Fig.~\ref{ZA}(a,~c), we found that the skyrmion number of  $\hat{\mathbf d}$ vector at magnon and ZA phonon crossing corresponds to $C_+=1$ at $\theta=\pi/8$ and corresponds to $C_+=-1$ at $\theta=3\pi/8$. In Fig.~\ref{TA}(a,~c), we found that the skyrmion number of  $\hat{\mathbf d}$ vector at magnon and TA phonon crossing corresponds to $C_+=2$ at $\theta=\pi/4$ and it corresponds to $C_+=-2$ at $\theta=3\pi/4$. 

\section{thermal Hall conductivity}
The finite Berry curvatures of magnon-phonon hybrid excitations give rise to the intrinsic thermal Hall effect as shown below. The semiclassical equations of motion for the wave packet of magnon-polarons are given by~\cite{xiao2010berry,sundaram1999wave}
\bea
\dot r=\frac 1\hbar \frac{\d E_n(\bk)}{\d\bk}-\dot{\bk}\times\Omega_n(\bk), \quad \hbar\dot \bk=-\nabla U(\r)\,,
\eea
where $U(r)$ is the potential acting on thewave packet which can be regarded as a confining potential of the bosonic excitation.

The Berry-curvature-induced thermal Hall conductivity is given by~\cite{matsumoto2011theoretical,matsumoto2011rotational}
\bea
\kappa^{xy}=-\frac{k^2_B T}{\hbar V}\sum_{n,\bk}c_2(\rho_{n,\bk})\Omega_n(\bk)\,,
\eea
where $c_2(\rho)=(1+\rho)\ln^2[(1+\rho)/\rho]-\ln^2\rho-2\text{Li}_2(-\rho)$, $\rho_{n,\bk}=[e^{E_n(\bk)/k_BT}-1]^{-1}$ is Bose-Einstein distribution function with a zero chemical potential, $k_B$ is the Boltzmann constant, T is the temperature, and $\text{Li}_2(z)$ is the polylogarithm function.

\begin{figure}[t]
	\centering
	\subfloat[]{\includegraphics[width=0.45\linewidth]{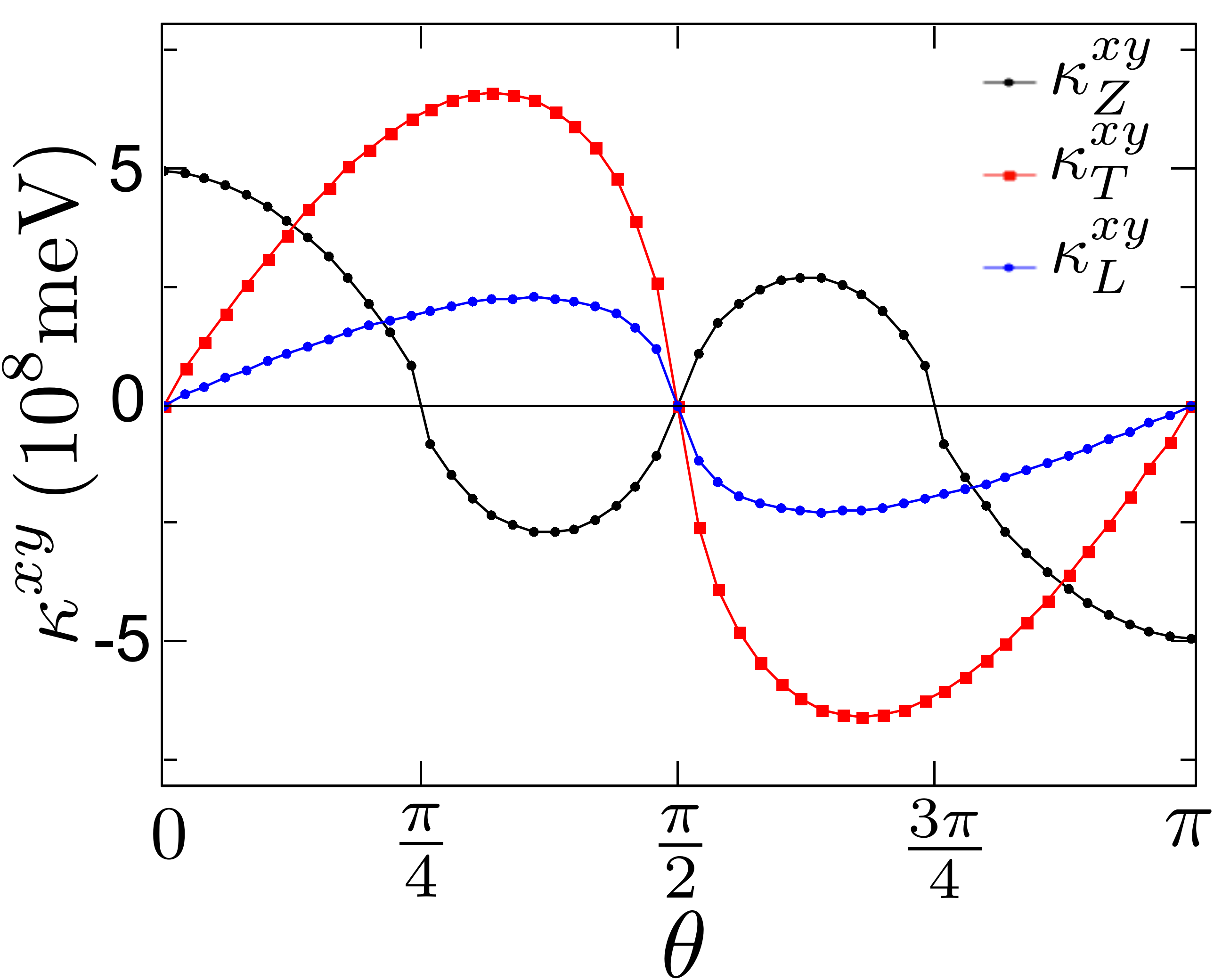}}
	\hfill
	\subfloat[]{\includegraphics[width=0.46\linewidth]{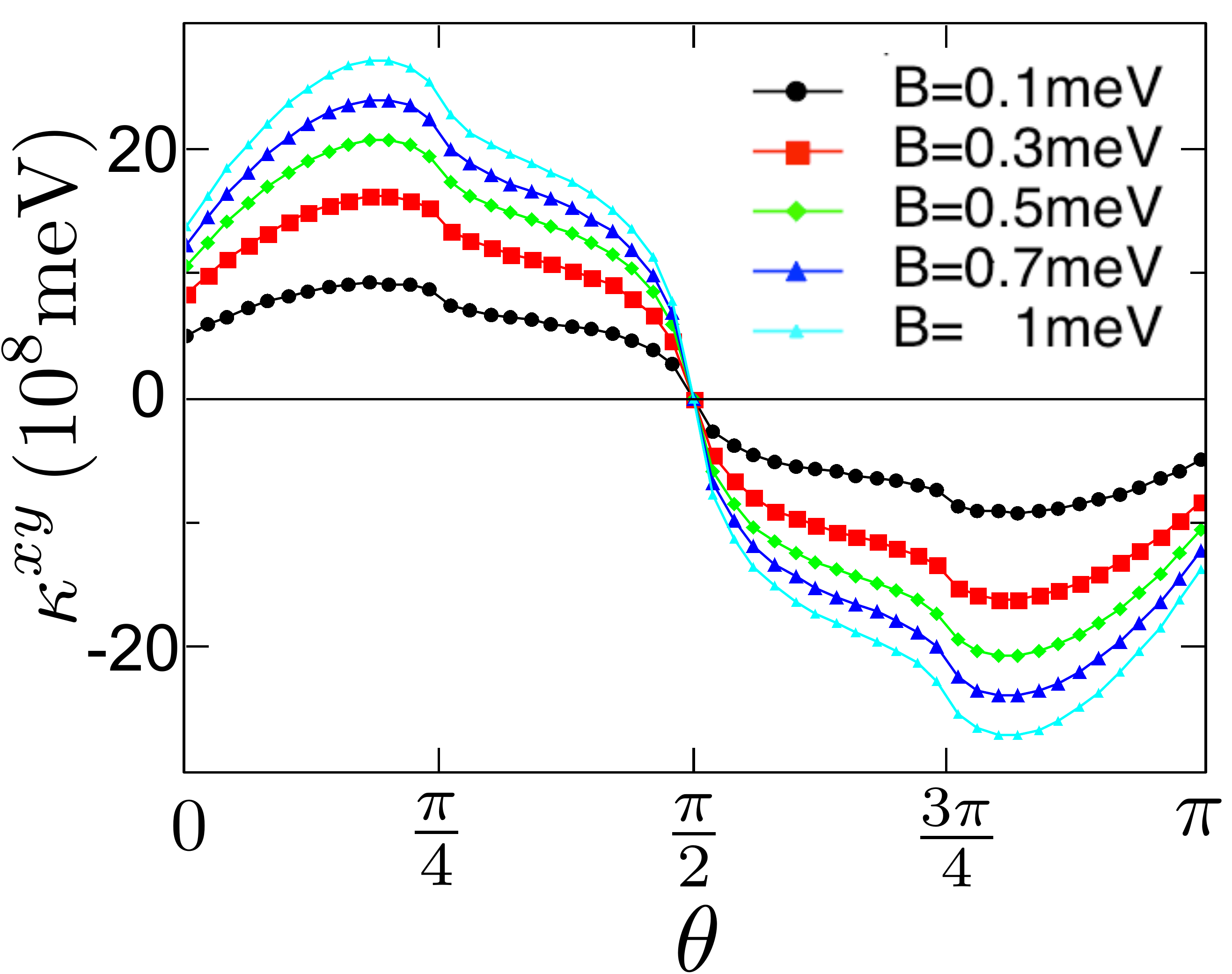}}
	\caption{(a) The contributions $\kappa_Z^{xy}, \kappa_T^{xy}$ and $\kappa_L^{xy}$ to the thermal Hall conductivities from magnon-ZA phonon coupling, magnon-LA phonon coupling, and magnon-TA phonon coupling, respectively. Parameter used are T=10K and B=0.1meV. (b) Dependence of the total thermal Hall conductivity $\kappa^{xy}$ on the tilt angle for several different magnetic fields at T=10\,K.}
	\label{kappa}
\end{figure}

In Fig.~\ref{kappa}(a), we show the dependence of thermal Hall conductivity on the magnetic field tilt angle from contribution of magnon with ZA, LA, TA phonon coupling separately at $B=0.1$~meV. At $\theta=0$, there is no contribution from LA and TA phonon because there is no Berry curvature from them. Across $\theta=\pi/4$, the contribution of ZA phonon becomes negative because the Berry curvature change sign at this angle. At $\theta=\pi/2$, when the ground state spin direction is in-plane, all the contributions vanish and the thermal Hall conductivity becomes zero as expected from the general symmetry grounds as follows. The thermal Hall conductivity is odd under time reversal: $\kappa^{xy}(\theta)=-\kappa^{xy}(\pi+\theta)$. Our system respects the two-fold rotational symmetry: $\kappa^{xy}(\theta)=\kappa^{xy}(2\pi-\theta)$. Then, it follows $\kappa^{xy}(\theta)=-\kappa^{xy}(\pi-\theta)$. In particular, we have $\kappa^{xy}(\pi/2)=0$, implying the absence of the thermal Hall effect for the in-plane magnetic field. In summary, the thermal Hall conductivity starts from a finite positive value at $\theta=0$. It starts to increase as the angle increases and reaches the maximum, then decreases to the negative minimum by passing zero at $\theta=\pi/2$. Eventually, it increases to a negative value at $\theta=\pi$ as shown in Fig.~\ref{kappa}(b). 

\section{discussion}
In this paper, we studied topological properties of magnon-polarons in a square lattice ferromagnet with Kittel's magnetoelastic interaction subjected to a magnetic field in an arbitrary direction. 
In our model, the magnons couples with longitudinal and transverse in-plane phonons as well as  out-of-plane phonons.  We investigated the topological structure of the magnon-polaron bands by mapping our model to the well-known two-band model of topological insulator near each band crossing point. The Berry curvature and the Chern number corresponding to three hybridizations are found to differ as follows. First, magnons and out-of-plane phonons coupling gives rise to a Chern number $\abs C$=1, which  changes sign at $\theta=\pi/4$, $\pi/2$ and $3\pi/4$.  Secondly, magnons and transverse in-plane phonons coupling gives rise to a Chern number $\abs C=2$, which changes sign at $\theta=\pi/2$. Thirdly, the hybridization of magnons and longitudinal in-plane phonons does not possess the well-defined Chern number since there are two points in the momentum space where the gap is not opened up. We also calculated the dependence of thermal Hall conductivity based on our model as an experimental probe. The unique behavior of thermal Hall conductivity as a function of the field direction reflects the contributions from in-plane and out-of plane phonons and can be used to probe the topological tunability of the magnon-polaron bands. 

Kittel's magnetoelastic interaction originates from the magnetic anisotropy, which is ubiquitous in ferromagnetic thin film structure~\cite{dieny2017perpendicular}, but the effect of magnetoelastic coupling on magnetic dynamics is still largely unknown. In our model, we used a Hamiltonian that is bilinear in the magnon and phonon operator, which result in the hybridization of magnon and phonon mode, by neglecting higher-order terms. If magnetoelastic interaction is sufficiently strong , the band will be strongly renormalized~\cite{furukawa1999magnon} and may require the inclusion of the higher-order terms. The physics of strong magnetoelastic interaction could be more pronounced in two dimensional magnets with weak mechanical stability~\cite{woods2001magnon, burch2018magnetism}, which will be pursued in the future.

\begin{acknowledgments}
The authors appreciate the useful discussions with Kyung-Jin Lee, Gyungchoon Go, and Shu Zhang. This work is supported by the University of Missouri. S.K.K. acknowledges Young Investigator Grant (YIG) from Korean-American Scientists and Engineers Association (KSEA). 
\end{acknowledgments}

\bibliography{bibfile}

\end{document}